\documentclass[11pt,twoside]{article}
\usepackage{asp2004}
\usepackage{psfig}
\usepackage{epsf}
\usepackage{graphics}
\usepackage{lscape}
\def\edcomment#1{\iffalse\marginpar{\raggedright\sl#1\/}\else\relax\fi}
\reversemarginpar

\begin{document}

\title{Virtual Observatory: From Concept to Implementation}

\author{S.G. Djorgovski$^{1,2}$ and R. Williams$^2$}

\affil{$^1$ Division of Physics, Mathematics, and Astronomy \\
       $^2$ Center for Advanced Computing Research \\
       California Institute of Technology \\
       Pasadena, CA 91125, USA}

\begin{abstract}
We review the origins of the Virtual Observatory (VO) concept, and the 
current status of the efforts in this field.  VO is the response of the
astronomical community to the challenges posed by the modern massive and
complex data sets.  It is a framework in which information technology
is harnessed to organize, maintain, and explore the rich information 
content of the exponentially growing data sets, and to enable a
qualitatively new science to be done with them.  VO will become a
complete, open, distributed, web-based framework for astronomy of the
early 21st century.  A number of significant efforts worldwide are now
striving to convert this vision into reality.  The technological and
methodological challenges posed by the information-rich astronomy are
also common to many other fields.  We see a fundamental change in the
way all science is done, driven by the information technology revolution.
\end{abstract}
\thispagestyle{plain}

\section{The Challenge and the Opportunity}

Like all other sciences, and indeed most fields of the modern human
endeavor (commerce, industry, security, entertainment, etc.), 
astronomy is being deluged by an exponential growth in the volume and
complexity of data.  The volume of information gathered in astronomy
is estimated to be doubling every 1.5 years or so (Szalay \& Gray 2000),
i.e., with the
same exponent as the Moore's law.  This is not an accident: the same
technology which Moore's law describes (roughly, VLSI) has also given
us most astronomical detectors (e.g., CCDs) and data systems.  The
current ($\sim$ early 2005) data gathering rate in astronomy is estimated
to be $\sim 1$ TB/day, and the content of astronomical archives is
now several hundred TB (Brunner et al. 2002).  Note that both the data
volume and the data
rate are growing exponentially.  Multi-PB data sets are on the horizon.

In addition to the growth in data volume, there has been also a great
increase in data complexity, and generally also quality and homogeneity.
The sky is now being surveyed at a full range of wavelengths, from
radio to $\gamma$-rays, with individual surveys producing data sets
measured in tens of TB, detecting many millions or even billions of
sources, and measuring tens or hundreds of parameters for each
source.  There is also a bewildering range of targeted observations,
many of which carve out multi-dimensional slices of the parameter
space.

Yet, our understanding of the universe is clearly not doubling every
year and a half.  It seems that we are not yet exploiting the full
information content of these remarkable (and expensive) data sets.
There is something of a technological and methodological bottleneck
in our path from bits to knowledge.

A lot of valuable data from ground-based observations is not yet
archived or documented properly.  A lot of data is hard to find
and access in practice, even if it is available in principle.
A multitude of good archives, data depositories, and digital 
libraries do exist, and form an indispensable part of the 
astronomical research environment today.  However, even those
functional archives are like an archipelago of isolated islands
in the web, which can be accessed individually one at a time,
and from which usually only modest amounts of data can be
downloaded to the user's machine where the analysis is done.

Even if one could download the existing multi-TB data sets in
their full glory (a process which would take a long, long time
even for the well-connected users), there are no data exploration
and analysis tools readily available, which would enable actual
science with these data to be done in a reasonable and practical
amount of time.

We are thus facing an embarrassment of richness: a situation where
we cannot effectively use the tremendous -- and ever growing --
amounts of valuable data already in hand.  Fortunately, this is
a problem where technological solutions do exist or can be
developed on a reasonably short time scale.

\section{The Genesis of the Virtual Observatory Concept}

The Virtual Observatory (VO) concept is the astronomical community's
response to these challenges and opportunities.  VO is an emerging,
open, web-based, distributed research environment for astronomy with
massive and complex data sets.  It assembles data archives and services,
as well as data exploration and analysis tools.  It is technology-enabled,
but science-driven, providing excellent opportunities for collaboration
between astronomers and computer science (CS) and IT professionals and
statisticians.  It is also an example of a new type of a scientific
organization, which is inherently distributed, inherently multidisciplinary,
with an unusually broad spectrum of contributors and users.

\begin{figure}[!ht]
\plotone{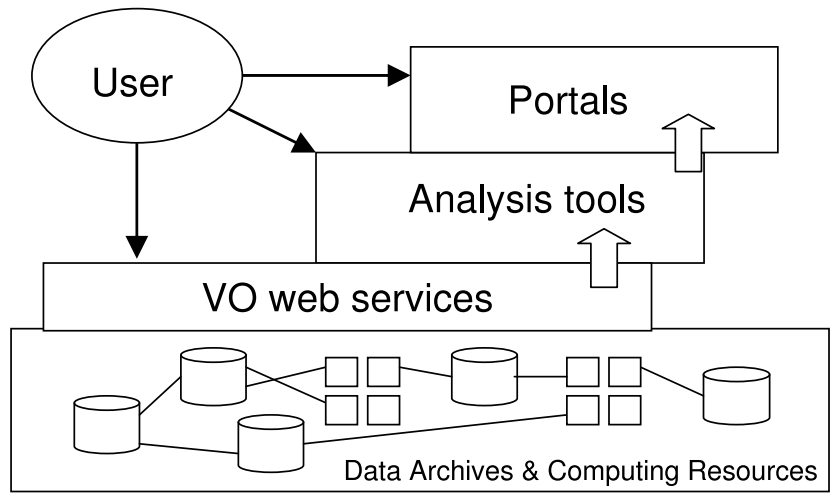}
\caption{
A Conceptual outline of a VO.  User communicates with a portal that provides 
data discovery,
access, and federation services, which operate on a set of interconnected
data archives and compute resources, available through standardized web services.  
User-selected or
generated data sets are then fed into a selection of data exploration
and analysis tools, in a way which should be seamlessly transparent
to the user.
}
\end{figure}

The concept was defined in the 1990's through many discussions and
workshops, e.g., during the IAU Symposium 179,
and at a special session at the 192nd meeting of the AAS
(Djorgovski \& Beichman 1998).  Precursor ideas include, e.g.,
efforts on the NASA's early ADS system, and its ESA counterpart,
ESIS, as well as the development of many significant data and
literature archives in the same period: ADS itself, Simbad and other
services at CDS, NED, data archives from the HST and other space
missions, Digital Sky project (by T. Prince et al.), etc.  As many
modern digital sky surveys (e.g., DPOSS,
SDSS, 2MASS, etc.) started producing terabytes of data, the challenges
and opportunities of information-rich astronomy became apparent
(see, e.g., Djorgovski et al. 1997, Szalay \& Brunner 1998, 
Williams et al. 1999, or Szalay et al. 2000).
Two grassroots workshops focused on the idea of VO were held in 1999,
at JHU (organized by A. Szalay, R. Hanisch, et al.), and at NOAO
(organized by D. De Young, S. Strom, et al.).

The early developments culminated in a significant endorsement of the
NVO concept by the U.S. National
Academy's ``astronomy decadal survey'' (McKee, Taylor, et al. 2000).
This was then explored further in a White Paper (2001), and in
other contributions to the first major conference on the subject
(Brunner, Djorgovski, \& Szalay 2001), from which emerged the architectural
concept of services, service descriptions, and a VO-provided registry.
The report of 
the U.S. National Virtual Observatory Science Definition Team (2002)
provided the most comprehensive scientific
description of the concept and the background up to that point.

More international conferences followed (e.g., Banday et al. 2001,
Quinn \& Gorski 2004), and a good picture of this
emerging field can be found in papers contained in their proceedings.
VO projects have been initiated world-wide, with a good and growing
international collaboration between various efforts;
more information and links can be found on their websites.
\footnote{ 
The U.S. National Virtual Observatory (NVO) project website: http://us-vo.org
}
\footnote{
The International Virtual Observatory Alliance (IVOA) website: http://ivoa.net
}

\begin{figure}[!ht]
\plotone{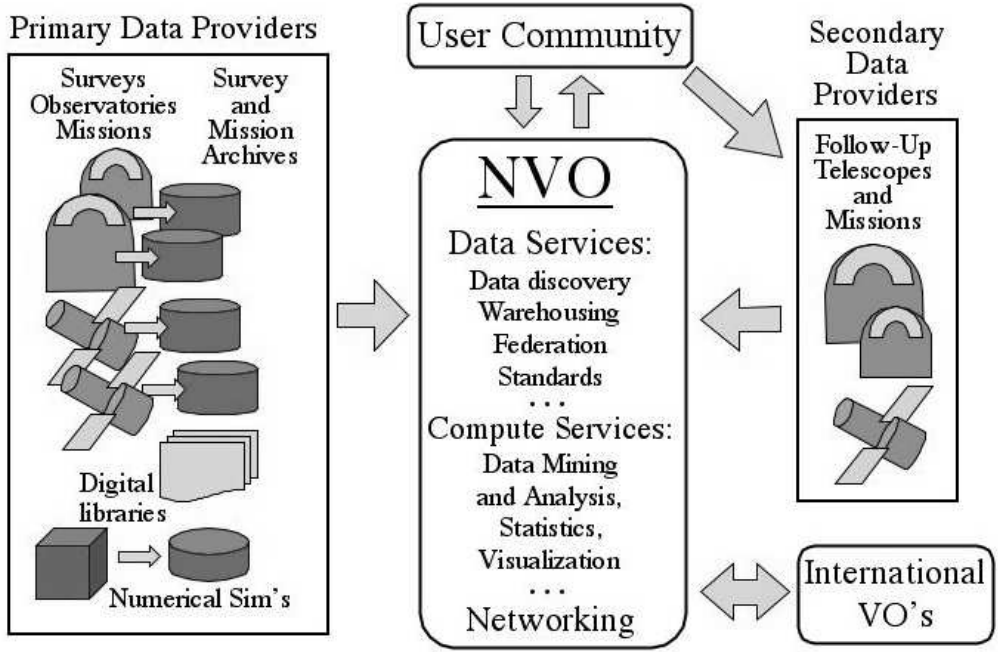}
\caption{
A systemic view of the VO as a complete astronomical research environment,
connecting archives of both ground-based and space-based observations,
and providing the tools for their federation and exploration.  Analysis
of archived observations -- some of which may be even real-time data --
then leads to follow-up observations, which themselves become available
within the VO matrix.
}
\end{figure}

Finally, VO can be seen as a connecting tissue of the entire astronomical
system of observatories, archives, and compute services (Fig. 2;
Djorgovski 2002).
Effectiveness of any observation is amplified, and the scientific
potential increased, as the new data are folded in the system and made
available for additional studies, follow-up observations, etc.

\section{Scientific Roles and Benefits of a VO}

The primary role of a VO is to facilitate data discovery (what is already
known of some object, set of objects, a region on the sky, etc.), data
access (in an easy and standardized fashion), and data federation (e.g.,
combining data from different surveys).  The next, and perhaps even more
important role, is to provide an effective set of data exploration and
analysis tools, which scale well to data volumes in a multi-TB regime,
and can deal with the enormous complexity present in the data.
(By ``data'' we mean both the products of observations, and products
of numerical simulations.)

While any individual function envisioned for the VO can be accomplished
using existing tools, e.g., federating a couple of massive data sets,
exploring them in a search for particular type of objects, or outliers, or
correlations, in most cases such studies would be too time-consuming and
impractical; and many scientists would have to solve the same issues
repeatedly.  VO would thus serve as an 
\emph{enabler of science with massive and complex data sets},
and as an 
\emph{efficiency amplifier}.  The goal is to enable
some qualitatively new and different science, and not just the same as
before, but with a larger quantity of data.  We will need to learn to ask
different kinds of questions, which we could not hope  to answer with the
much smaller and information-poor data sets in the past.

Looking back at the history of astronomy we can see that technological
revolutions lead to bursts of scientific growth and discovery. For
example, in the 1960's, we saw the rise of radio astronomy, powered by the
developments in electronics (which were much accelerated by the radar
technology of the World War II and the cold war).  This has led to the
discovery of quasars and other powerful active galactic nuclei, pulsars,
the cosmic microwave background (which firmly established the Big Bang
cosmology), etc.  At the same time, the access to space opened the fields
of X-ray and $\gamma$-ray astronomy, with an equally impressive range of
fundamental new discoveries:
the very existence of  the cosmic X-ray sources and the cosmic X-ray
background, $\gamma$-ray bursts (GRBs), and other energetic phenomena.
Then, over
the past 15 years or so, we saw a great progress powered by the advent of
solid-state detectors (CCDs, IR arrays, bolometers, etc.), and cheap and
ubiquitous computing, with discoveries of extrasolar planets, brown dwarfs,
young and forming galaxies at high redshifts, the cosmic acceleration (the
dark energy), the solution of the mystery of GRBs, and so on.  We are now
witnessing the next phase of the IT revolution, which will likely lead to
another golden age of discovery in astronomy.  VO is the framework to
effect this process.

In astronomy, observational discoveries are
usually made either by opening a new domain of the parameter space (e.g.,
radio astronomy, X-ray astronomy, etc.), by pushing further along some axis
of the observable parameter space (e.g., deeper in flux, higher in angular
or temporal resolution, etc.), by expanding the coverage of the parameter
space and thus finding rare types of objects or phenomena which would be
missed in sparse observations, or by making connections between different
types of observations (for example, optical identification of radio sources
leading to the discovery of quasars).  Surveys are often a venue which
leads to such discoveries; see, e.g., Harwit (1998) for a discussion.
In a more steady mode of research,
application of well understood physics, constrained by observations, leads
to understanding of various astronomical objects and phenomena; e.g.,
stellar structure and evolution.

This implies two kinds of discovery strategies: covering a large volume of
the parameter space, with many sources, measurements, etc., as is done very
well by massive sky surveys; and connecting as many different types of
observations as possible (e.g., in a multi-wavelength, multi-epoch, or
multi-scale manner), so that the potential for discovery increases as the
number of connections, i.e., as the number of the federated data sets,
squared.  Both approaches are naturally suited for the VO.

\section{Technological and Methodological Challenges}

There are many non-trivial technological and methodological problems posed
by the challenges of data abundance.  We note two important trends, which 
seem to particularly distinguish the new, information-rich science from 
the past:

\begin{itemize}

\item {\it Most data will never be seen by humans.}
This is a novel experience 
for scientists, but the sheer volume of TB-scale data sets (or larger) 
makes it impractical to do even a most cursory examination of all data.  
This implies a need for reliable data storage, networking, and 
database-related technologies, standars, and protocols.

\item {\it Most data and data constructs, and patterns present in them, 
cannot be comprehended by humans directly.}
This is a direct consequence of a 
growth in complexity of information, mainly its multidimensionality.  
This requires the use or development of novel data mining (DM) or 
knowledge discovery in databases (KDD) and data understanding (DU) 
technologies, hyperdimensional visialization, etc.  The use of 
AI/machine-assisted discovery may become a standard scientific practice.

\end{itemize}

This is where the qualitative differences in the way science is done in 
the 21st century will come from; the changes are not just quantitative, 
based on the data volumes alone.
Thus, a modern scientific discovery process can be outlined as follows:

\begin{enumerate}

\item {\bf Data gathering:} raw data streams produced by various measuring devices.  
Instrumental effects are removed and calibrations applied in the 
domain-specific manner, usually through some data reduction pipeline (DRP).

\item {\bf Data farming:}  storage and archiving of the raw and processed data, 
metadata, and derived data products, including issues of optimal database
architectures, indexing, searchability, interoperability, data fusion, etc.
While much remains to be done, these challenges seem to be fairly well
understood, and much progress is being made.

\item {\bf Data mining:}  including clustering analysis, automated classification,
outlier or anomaly searches, pattern recognition, multivariate correlation
searches, and scientific visualization, all of them usually in some
high-dimensional parameter space of measured attributes or imagery.
This is where the key technical challenges are now.

\item {\bf Data understanding:} converting the analysis results into the 
actual knowledge.  The problems here are essentially methodological in nature.
We need to learn how to ask new types of questions, enabled by the increases
in the data volume, complexity, and quality, and the advances provided by IT.
This is where the scientific creativity comes in. 

\end{enumerate}

For example, a typical VO experiment may involve federation of several major
digital sky surveys (in the catalog domain), over some large area of the
sky.  Each survey may contain $\sim 10^8 - 10^9$ sources, and measure
$\sim 10^2$ attributes for each source (various fluxes, size and shape
parameters, flags, etc.).  Each input catalog would have its own limits
and systematics.  The resulting data set would be somewhat heterogeneous
parameter space of $N \sim 10^9$ data vectors in $D \sim 10^2 - 10^3$
dimensions.  An exploration of such a data set may require a clustering
analysis (e.g., how many different types of objects are there? which
object belongs to which class, with what probability?), a search for
outliers (are there rare or unusual objects which do not belong to any
of the major classes?), a search for multivariate correlations (which may
connect only some subsets of measured parameters), etc.  For some
examples and discussion, see, e.g., Djorgovski et al. (2001ab, 2002).

The primary challenge is posed by the large size of data volume,
and -- especially -- large dimensionality.  The existing clustering
analysis algorithms do not scale well with the data volume $N$,
or dimensionality $D$.  At best, the processing time is proportional
to $N \log N$, but for some methods it can be $\sim N^\alpha$,
where $\alpha \geq 2$.  The curse of hyperdimensionality is even
worse, with typical scaling as $\sim D^\beta$, where $\beta \geq 2$;
most off-the-shelf applications can deal with $D < 10$.  Thus, the
computational cost itself may be prohibitive, and novel approaches
and algorithms must be developed.

In addition, there are many possible complications: data heterogeneity,
different flux limits, errors which depend on other quantities, etc.
In the parameter space itself, clustering may not be well represented by
multivariate Gaussian clouds (a standard approach), and distributions
can have power-law or exponential tails (this greatly complicates the
search for anomalies and rare events); and so on. As we search for outliers 
in these new rich surveys, the requirement to eliminate
noise and artifacts grows. These are much more significant in an outlier
search than when computing clustering properties and averages,

The difficulties of data federation are exacerbated if data is in different
formats and delivery mechanisms, increasing both manual labor and the
possibility of mistakes. Data with inadequate metadata description can
be misleading -- for example mistaking different equinoxes for proper
motion because the equinox was not stated.  Another difficulty can be
that delivery of the data is optimized for either browsing or for bulk
access: it is difficult if the user wants one, but the other is the only
option.  The Virtual Observatory has already provided many well-adopted
standards that were built for data fedteration.  The VOTable standard,
for example (Ochsenbein, Williams, et al. 2001) carries rich metadata
about tables, groups of tables, and the data dictionaries.

The second, and perhaps even more critical part of the curse of
hyperdimensionality is the visualization of these highly-dimensional data
parameter spaces.  Humans are biologically limited to visualize patterns and
scenes in 2 or 3 dimensions, and while some clever tricks have been developed
to increase the maximum visualizable dimensions, in practice it is hard to
push much beyond $D$ = 4 or 5.  Mathematically, we understand the meaning of
clustering and correlations in an arbitrary number of parameter space
dimensions, but how can we actually visualize such structures?  Yet,
recognizing and comprehending such complex data constructs may lead to
some crucial new astrophysical insights.  This is an essential part of the
intuitive process of scientific discovery, and critical to data understanding
and knowledge extraction.
Effective and powerful data visualization, applied in the parameter space
itself, must be an essential part of the interactive clustering analysis.

In many situations, scientifically informed input is needed in designing
and applying the
clustering algorithms. This should be based on a close, working collaboration
between astronomers and computer scientists and statisticians. There are too
many unspoken assumptions, historical background knowledge specific to 
astronomy, and opaque jargon; constant communication and interchange
of ideas are essential.

\section{The Virtual Observatory Implemented}

The objective of the Virtual Observatory is to improve and unify access
to astronomical data and services for primarily professional astronomers, 
but also for the general public and students. Figure 3 gives an overview. 
The top bar of the figure represents this objective: discovery of data and 
services, reframing and analysing that data through computation, publishing 
and dissemination of results, and increasing scientific output through 
collaboration and federation. The IVOA does not specify or recommend any 
specific portal or library by which users can access VO data, but some 
examples of these portals and tools are shown in the grey box. 

\begin{figure}[!ht]
\plotone{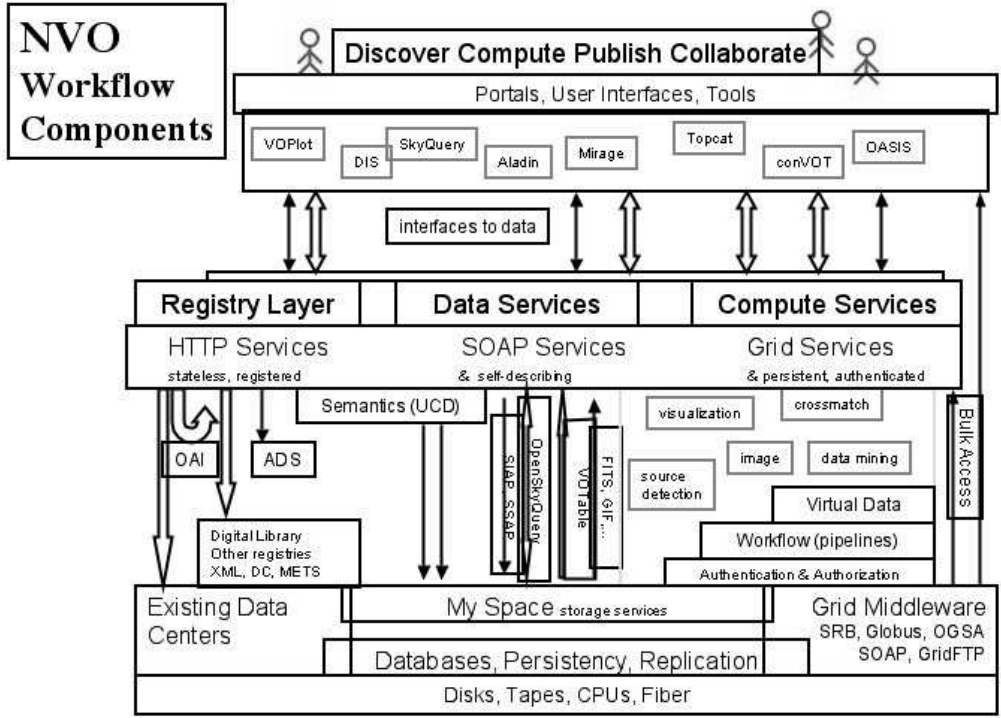}
\caption{
Internationally adopted architecture for the VO. Services are split into 
three kinds: fetching data, computing services, and registry (publishing 
and discovery). Services are implemented in simple way (web forms) and as 
sophisticated SOAP services. The VO does not recommend or endorse a 
particular portal for users, but rather encourages variety.
}
\end{figure}

Different vertical arrows represent the different service types and XML 
formats by which these portals interface to the IVOA-compliant services. 
In the IVOA architecture, we have divided the available services into 
three broad classes: 

\begin{enumerate}
\item Data Services, for relatively simple services that provide access 
to data.
\item Compute Services, where the emphasis is on computation and federation 
of data. 
\item Registry Services, to allow services and other entities to be 
published and discovered.
\end{enumerate}

These services are implemented at various levels of sophistication, from 
a stateless, text-based request-response, up to an authenticated, 
self-describing service that uses high-performance computing to build a 
structured response from a structured request. In the VO, it is intended 
that services can be used not just individually, but also concatenated in 
a distributed workflow, where the output of one is the input of another. 

The registry services facilitate publication and discovery of services. 
If a data center (or individual) puts a new dataset online, with a service 
to provide access to it, the next step would be to publish that fact to a 
VO-compliant registry. One way to do this is to fill in forms expressing 
who, where, and how for the service. In due course, registries harvest 
each other (copy new records) and so the new dataset service will be 
known to other VO-registries. When another person searches a registry 
(by keyword, author, sky region, wavelength, etc), they will discover 
the published services. In this way the VO advances information diffusion 
to a more efficient and egalitarian system. 

In the VO architecture, there is nobody deciding what is good data and 
what is bad data, (although individual registries may impose such criteria 
if they wish). Instead, we expect that good data will rise to prominence 
organically, as it does on the World Wide Web. We note that while the web 
has no publishing restrictions, it is still an enormously useful resource; 
and we hope the same paradigm will make the VO registries useful. 

Each registry has three kinds of interface: publish, query, and harvest. 
People can publish to a registry by filling in web forms in a web portal, 
thereby defining services, data collections, projects, organizations, and 
other entities. The registry may also accept queries in a one or more 
languages (for example an IVOA standard Query Language), and thereby 
discover entities that satisfy the specified criteria. The third interface, 
harvesting, allows registries to exchange information between themselves, 
so that a query executed at one registry may discover a resource that was 
published at another. 

Registry services expect to label each VO resource through a universal 
identifier, that can be recognized by the initial string ivo://. 
Resources can contain links to related resources, as well as external links 
to the literature, especially to the Astronomical Data System. The IVOA 
registry architecture is compliant with digital library standards for 
metadata harvesting and metadata schema, with the intention that 
IVOA-compliant resources can appear as part of every University library. 

Data services range from simple to sophisticated, and return tabular, image, 
or other data. At the simplest level (conesearch), the request is a cone on 
the sky (direction/angular radius), and the response is a list of ``objects''
each of which has a position that is within the cone. Similar services (SIAP, 
SSAP) can return images and spectra associated with sky regions, and these 
services may also be able to query on other parameters of the objects. 

The OpenSkyQuery protocol drives a data service that allows querying of a 
relational database or a federation of databases. In this case, the request 
is written in a specific XML abstraction of SQL that is part of ADQL 
(Astronomical Data Query Language). 

The IVOA architecture will also support queries written at a more semantic 
level, including queries to the registry and through data services. To 
achieve this, the IVOA is developing a structured vocabulary called UCD 
(Unified Content Descriptor) to define the semantic type of a quantity. 

The IVOA expects to develop standards for more sophisticated services, for 
example for federating and mining catalogs, image processing and source 
detection, spectral analysis, and visualization of complex datasets. These 
services will be implemented in terms of industry-standard mechanisms, 
working in collaboration with the grid community. 

Members of the IVOA are collaborating with a number of IT groups that are 
developing workflow software, meaning a linked set of distributed services 
with a dataflow paradigm. The objective is to reuse component services to 
build complex applications, where the services are insulated from each other 
through well-defined protocols, and therefore easier to maintain and debug. 
IVOA members also expect to use such workflows in the context of Virtual 
Data, meaning a data product that is dynamically generated only when it is 
needed, and yet a cache of precomputed data can be used when relevant. 

In the diagram above, the lowest layer is the actual hardware, but above 
that are the existing data centers, who implement and/or deploy IVOA standard 
services. Grid middleware is used for high-performance computing, data 
transfer, authentication, and service environments. Other software components 
include relational databases, services to replicate frequently used 
collections, and data grids to manage distributed collections. 

A vital part of the IVOA architecture is {\it VOSpace} so that users can 
store data within the VO. VOSpace stores files and DB tables on the greater 
internet, and has a good security model so that legitimate data is secure 
and illegitimate data disallowed.  VOSpace avoids the need to recover results
to the desktop for storage or to keep them inside the service that generated 
them. Using VOSpace establishes access rights and privacy over intermediate 
results and allows users to manage their storage remotely.

\section{Examples of Some Prototype Services}

There are several deployed applications available at the NVO web site
\footnote{ 
NVO Applications: http://us-vo.org/apps
}. 
A registry portal allows the user to find source catalogs, image and 
spectral services, data sets and other astronomical resources registered 
with the NVO. 
OpenSkyQuery provides sophisticated selection and cross-match services 
from uploaded (user) data with numerous catalogues.
There are spectrum services for search, plot, and retrieving SDSS, 2dF, 
and other archives. 
The WESIX service asks the user to upload an image to a source-extraction 
code, then cross-correlates the objects found with selected survey catalogs. 
There is also information about publishing to the NVO and what that means.
 
As noted above, these are all implemented with Web Services. This means 
that users can effectively scale up their usage of NVO services: when 
the user finds the utility of a remote service that can be used by 
clicking on a human-oriented web page, there is often a further 
requirement to scale up by scripting the usage -- a machine-oriented 
interface. We have ensured in the VO architecture that there is always 
a straightforward programming interface behind web page.

In the following, we examine in more detail the DataScope service.
\begin{figure}[!ht]
\plotone{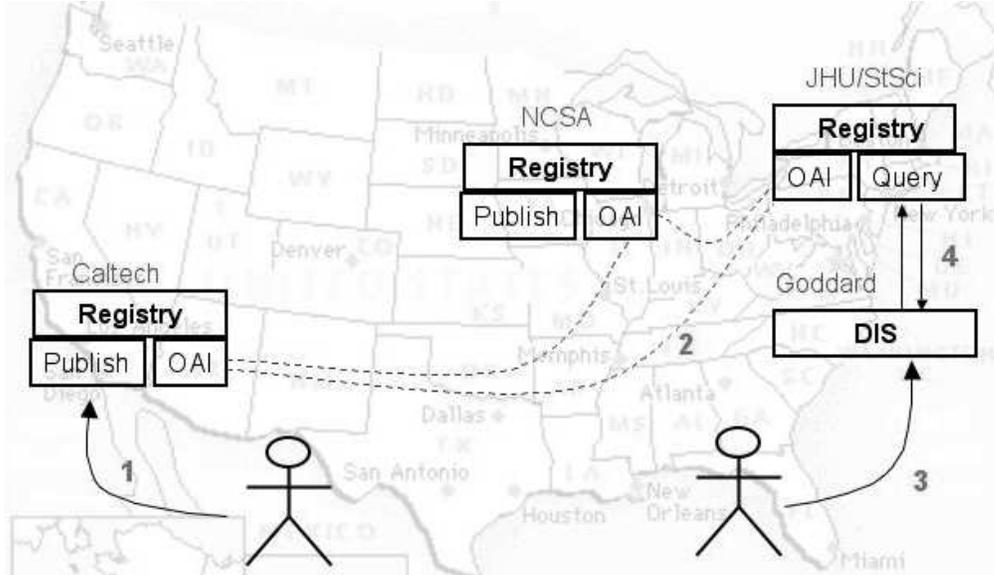}
\caption{
The DataScope service ``publish and discover'' paradigm. After a new data 
resource is published to a VO-compliant registry (1), the different 
registries harvest each other (2). When a query comes to DataScope (3), 
the new resource can be seen in federation with others (4). 
}
\end{figure}

Using NVO DataScope scientists can discover and explore hundreds of data 
resources available in the Virtual Observatory. Users can immediately 
discover what is known about a given region of the sky: they can view 
survey images from the radio through the X-ray, explore observations 
from multiple archives, find recent articles describing analysis of 
data in the region, find known interesting or peculiar objects and 
survey datasets that cover the region. There is a summary of all of 
the available data. Users can download images and tables for further 
analysis on their local machines, or they can go directly to a growing 
set of VO enabled analysis tools, including Aladin, OASIS, VOPlot and 
VOStat. 

As illustrated in Figure 4, DataScope provides a dynamic, simple to use 
explorer for VO data, protocols and analysis tools. Developed by Tom 
McGlynn at NASA/GSFC and collaborators at STScI and NCSA, the DataScope 
uses the distributed VO registry and VO access protocols to link to 
archives and catalogs around the world.

There are web sites that provide rapid collection and federation of 
multi-wavelength imaging, catalog and observation data. This sort of 
interface has been built before (NED, Astrobrowse, Skyview, VirtualSky, 
etc). In these excellent and competent systems, data may be harvested and 
processed in advance, and there may be a lot of effort by devoted human 
curators. There may be links to remote resources -- but no guarantee 
that anything will actually be found ``under the link''.

However, the DataScope is different from these web sites for two major
reasons. First, the ``Publish and Discover'' paradigm means that 
DataScope is always up-to-date. When the sky position is given by the 
user, DataScope probes a collection of services to get relevant data, 
and that collection is fetched dynamically by querying the NVO Resource 
Registry. Therefore, when a new data service is created and published to 
the Registry, that service is immediately visible to the scientific 
community as part of Datascope.

Second, the DataScope uses standards. The NVO has defined standard service 
types for querying catalog and image servers. This replaces the old system, 
where each service implementor would choose an idiosyncratic interface, 
meaning that the maker of a federation service would need to learn and 
program each data service individually.

\section
{Taking a Broader View: Information-Intensive Science for the 21st Century}

The modern scientific methodology originated in the 17th century, and a
healthy interplay of analytical and experimental work has been driving the
scientific progress ever since.  But then, in mid-20th century, something
new came along: computing as a new way of doing science, primarily through
numerical simulations of phenomena too complex to be analytically tractable.
Simulations are thus more than just a substitute for analytical theory:
there are many phenomena in the physical universe where simulations
(incorporating, of course, the right physics and equations of motion) are
the only way in which some phenomena can be described and predicted.  Recall
that even the simplest Newtonian mechanics can solve exactly only a 2-body
problem; for $N>3$, numerical solutions are necessary.  Other examples in
astronomy include star and galaxy formation, dynamics and evolution of
galaxies and large-scale structure, stellar explosions, anything involving
turbulence, etc.  Simulations relate, can stimulate, or be explained by
both analytical theory and experiments or observations. While numerical
simulations and other computational means of solving complex systems of
equations continue to thrive, there is now a new and growing role of
scientific computing, which is data-driven.

Data- or information-driven computing, which spans all of the aspects 
of a modern scientific work described above, and more, is now becoming 
the dominant form of scientific computing, and an essential component 
of gathering, storing, preserving, accessing, and, most of all, analyzing
massive amounts of complex data, and extracting knowledge from them.  
It is fundamentally changing the way in which science is done in the 21st
century.

Computationally driven and enabled science also plays an important societal
role: it is empowering an unprecedented pool of talent.  With distributed
scientific frameworks like VO, which provide open access to data and tools
for their exploration, anyone, anywhere, with a decent internet connection
can do a first rate science, learn about what others area doing, and
communicate their results.  This should be a major boon for countries
without expensive scientific facilities, and individuals at small or
isolated institutions.  The human talent is distributed geographically
much more broadly than money or other resources.

\section{Concluding Comments}

The VO concept is rapidly spreading in astronomical community worldwide.
Ultimately, it should become ``invisible'', and taken for granted:
it would be $the$ operating framework for astronomical research,
a semantic web of astronomy.

There is an already effective, world-wide collaboration between various
national and trans-national VO efforts in place.  The fundamental
cyber-infra\-struc\-ture of interoperable data archives, standard formats,
protocols, etc., and a number of useful prototype services are well
under way.  The next stage of technological challenges is in the broad
area of data exploration and understanding (DM/KDD/DU).  We are
confident that continuing productive collaborations among astronomers,
statisticians, and CS/IT scientists and professionals will bring forth
a powerful new toolbox for astronomy with massive and complex data sets.

Just as technology derives from a progress in science, progress in 
science, especially experimental/observational, is driven by the progress 
in technology.  This positive feedback loop will continue, as the IT
revolution unfolds. Practical CS/IT solutions cannot be developed in a
vacuum; having real-life testbeds, and functionality driven by specific 
application demands is essential. Recall that the WWW originated as a
scientific application. Today, grid technology is being developed by 
physicists, astronomers, and other scientists.  The needs of 
information-driven science are broadly applicable to 
information-intensive economy in general, as well as other domains
(entertainment, media, security, education, etc.).  Who knows what
world-changing technology, perhaps even on par with the WWW itself,
would emerge from the synergy of computationally enabled science,
and science-driven information technology?

{\bfseries Acknowledgments.}
We thank the numerous friends and collaborators who developed the ideas
behind the Virtual Observatory, and made it a reality; they include
Charles Alcock,
Robert Brunner,
Dave De Young,
Francoise Genova,
Jim Gray,
Bob Hanisch, 
George Helou,
Wil O'Mullane,
Ray Plante,
Tom Prince,
Arnold Rots,
Alex Szalay, 
and many, many others, too numerous to list
here, for which we apologize.  We also acknowledge a partial support
from the NSF grants AST-0122449, AST-0326524, AST-0407448, and 
DMS-0101360, NASA contract NAG5-9482, and the Ajax Foundation.

\end{document}